\documentclass[12pt]{article}
\usepackage{a4wide,epsfig,psfrag,amsmath,amssymb,cite,scalefnt}
\usepackage{color}

\graphicspath{ {figs_pap/} }

\newcommand{\num}[1]{\color{black} #1}

\allowdisplaybreaks

\newcommand{\gsim}{\;\rlap{\lower 3.5 pt \hbox{$\mathchar \sim$}} \raise 1pt
 \hbox {$>$}\;}
\newcommand{\lsim}{\;\rlap{\lower 3.5 pt \hbox{$\mathchar \sim$}} \raise 1pt
 \hbox {$<$}\;}

\begin{document}

\title{\vskip-3cm{\baselineskip14pt
    \begin{flushleft}
      \normalsize TTP16-051
  \end{flushleft}}
  \vskip1.5cm
  On top quark mass effects to $gg\to ZH$ at NLO 
  \\[1em]
}

\author{
  Alexander Hasselhuhn,
  Thomas Luthe,
  Matthias Steinhauser
  \\[1em]
  {\small\it Institut f{\"u}r Theoretische Teilchenphysik}\\
  {\small\it Karlsruhe Institute of Technology (KIT)}\\
  {\small\it 76128 Karlsruhe, Germany}  
}
  
\date{}

\maketitle

\thispagestyle{empty}

\begin{abstract}
  We compute next-to-leading order QCD corrections to the process $gg\to ZH$.
  In the effective-theory approach we confirm the results in the
  literature. We consider top quark mass corrections via an asymptotic
  expansion and show that there is a good convergence
  below the top quark threshold which describes approximately {\num a quarter}
  of the total cross section.  Our corrections are implemented in
  the publicly available {\tt C++} program {\tt ggzh}.

%

\end{abstract}

\thispagestyle{empty}




\section{Introduction}

In the upcoming years the general purpose experiments ATLAS and CMS at the
CERN LHC will collect a large amount of data which will be used to perform
precision studies of various quantities. Among them are certainly the
properties of the Higgs boson, in particular its couplings to the other
particles of the Standard Model. Important quantities in this context are the
production cross sections and partial decay rates of the Higgs boson.
The dominant production process is via gluon fusion
followed by vector boson fusion and the so-called Higgs-strahlung process
$pp\to VH$ ($V=Z,W$) which is the subject of the current paper. 
Although $pp\to VH$ has a much smaller cross section it is a promising
channel to observe, e.g., if the Higgs boson decays to a
$b\bar{b}$ pair once 
substructure techniques are applied~\cite{Butterworth:2008iy}.

The leading order (LO) cross section is obtained from the Drell-Yan process
for the production of a virtual gauge boson $V^\star$ and its subsequent decay
into $VH$. Next-to-next-to-leading order QCD corrections to this channel have
been computed in
Refs.~\cite{Brein:2003wg,Brein:2011vx,Ferrera:2011bk,Ferrera:2013yga,Ferrera:2014lca}
and electroweak corrections have been considered in
Refs.~\cite{Ciccolini:2003jy,Denner:2011id}.
QCD corrections up to NNLO and electroweak corrections up to NLO for the total cross
section have been implemented in the program {\tt vh@nnlo}~\cite{Brein:2012ne}.

In Ref.~\cite{Kniehl:1990iva} the loop-induced production channel $gg\to ZH$
has been computed at leading order. NLO QCD corrections have been
computed in Ref.~\cite{Altenkamp:2012sx} in the heavy
top quark limit which significantly simplifies the
calculation. They are also implemented in {\tt vh@nnlo}~\cite{Brein:2012ne}.
Note that the NLO corrections to $gg\to ZH$ are formally N$^3$LO contributions
to $pp\to ZH$. However, due to the numerical importance of the gluon-induced 
process it is worthwhile to compute $gg\to ZH$ to NLO accuracy.

In this paper we study the effect of a finite top quark mass. At LO exact
results are available. However, at NLO the occurring integrals are highly
nontrivial and their evaluation is beyond straightforward application of
current multi-loop techniques.  We investigate the mass effects by expanding
the amplitudes for large $m_t$. This approximation is not valid in all phase
space regions. However, it provides an estimate of the numerical size of the
power-suppressed terms and thus of the quality of the effective-theory result.
Furthermore, it constitutes an important reference for a future exact
result since we observe a good convergence of the partonic cross sections
below the top quark pair threshold.  We only consider the $gg$ channel;
similar techniques can also be applied to the loop-induced contributions
of the $qg$ and $q\bar{q}$ channels which are, however, numerically much
smaller~\cite{Altenkamp:2012sx}. In our calculation we do not
consider decays of the final-state $Z$ boson.

Similar to $gg\to ZH$ also the process $gg\to HH$ is mediated by heavy quark
loops. NLO and NNLO corrections have been considered in a series of
papers~\cite{Dawson:1998py,Grigo:2013rya,deFlorian:2013jea,Shao:2013bz,Maltoni:2014eza,Grigo:2014jma,Grigo:2015dia,deFlorian:2015moa,Degrassi:2016vss,deFlorian:2016uhr}
applying various approximations. Recently the exact NLO corrections became
available~\cite{Borowka:2016ehy,Borowka:2016ypz}. The comparison to the
approximations shows sizeable differences for the total cross section and the
Higgs transverse momentum distribution. However, reasonable agreement between
the exact and the in $1/m_t$-expanded results is found for the Higgs pair
invariant mass ($m_{HH}$) distribution for not too large values of $m_{HH}$ if
the approximated result is re-scaled with the exact LO cross section.  Note
that the region between the production threshold and the top quark threshold
corresponds to about 100~GeV in the case of $HH$ and to about 135~GeV in the
case of $ZH$ production which makes the heavy-top expansion more interesting
for the latter.

Top quark mass effects have also been computed for the related process $gg\to
ZZ$. In Ref.~\cite{Melnikov:2015laa} $1/m_t^2$ corrections have been computed
at NLO, and interference effects
have been considered in~\cite{Campbell:2016ivq}.
In the latter reference Pad\'e approximation and conformal mapping has been
applied to improve the validity of the expansion in $1/m_t$.

The remainder of the paper is organized as follows: In Section~\ref{sec::LO}
we briefly discuss the LO cross section and compare the in $1/m_t$ expanded
and exact results. In Section~\ref{sec::part} we present our findings for the
partonic NLO cross section. In particular, we identify the approximation
procedure which leads to promising hadronic results, subject of
Section~\ref{sec::hadr_nlo}. We summarize our results in
Section~\ref{sec::concl}.


\section{\label{sec::LO}$gg\to ZH$ at LO}

\begin{figure}[t]
  \centering
  \begin{tabular}{cccc}
  \includegraphics[width=0.2\textwidth]{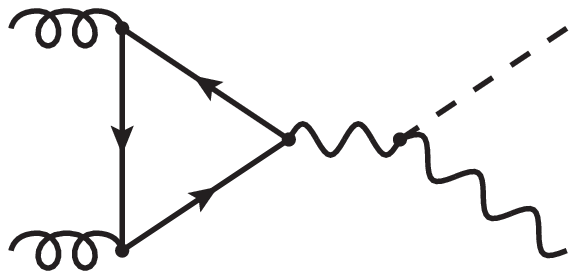} &
  \includegraphics[width=0.2\textwidth]{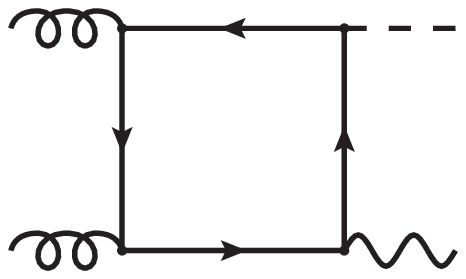} &
  \includegraphics[width=0.2\textwidth]{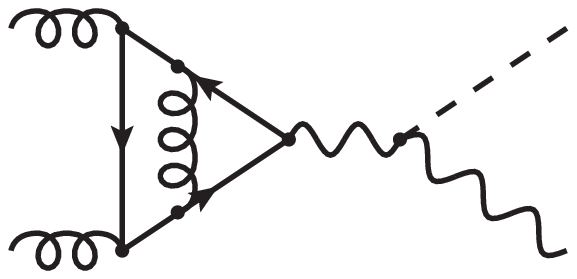} &
  \includegraphics[width=0.2\textwidth]{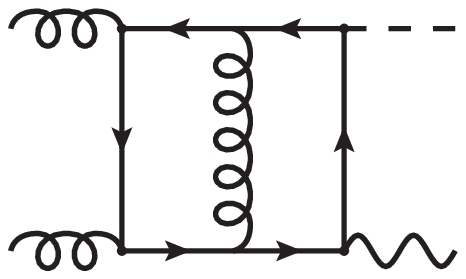}
  \\
  (a)&(b)&(c)&(d)
  \end{tabular}
  \\
  \begin{tabular}{ccc}
  \includegraphics[width=0.1\textwidth]{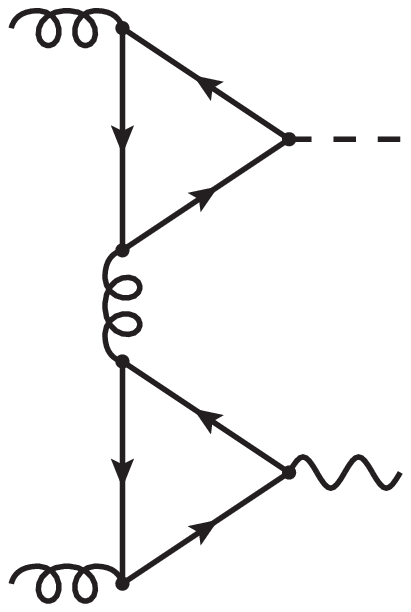}&
  \includegraphics[width=0.2\textwidth]{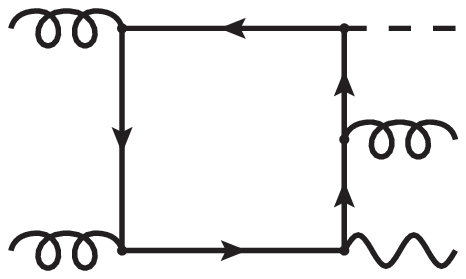}&
  \includegraphics[width=0.2\textwidth]{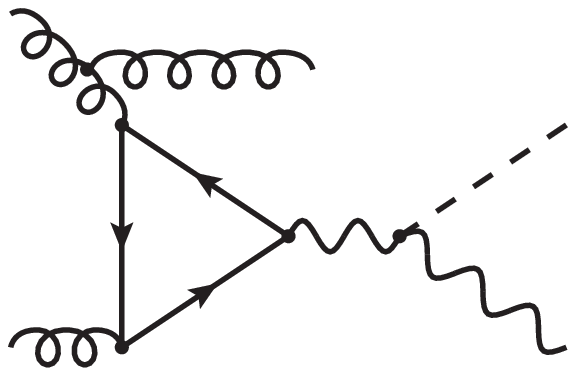}
  \\
  (e)&(f)&(g)
  \end{tabular}
  \caption{\label{fig::diag}Sample Feynman diagram contributing to $gg\to ZH$
    at LO and NLO. Solid, wavy, dashed and curly lines denote quarks, $Z$
    and Higgs bosons, and gluons, respectively. Internal wavy lines
    can also represent Goldstone bosons.}
\end{figure}

Sample Feynman diagrams contributing to the LO cross section are shown in
Fig.~\ref{fig::diag} (a) and (b). There are triangle contributions where the final-state
$Z$ and Higgs bosons are produced via a $s$-channel $Z$ or $\chi$ boson
exchange.  Both bottom and top quarks can be present in the loop.  In the case
of the box diagrams the Higgs boson couples directly to the quark running in
the loop and thus only internal top quarks are present since we neglect the
bottom Yukawa coupling. The effect of a finite bottom quark mass on the LO
cross section is at the per mille level.

In the heavy-$m_t$ approximation the diagrams with internal top quarks reduce
to vacuum integrals. The massless triangle diagrams are computed with the help
of simple form factor-type integrals which can be expressed in terms on
$\Gamma$ functions (see, e.g., Appendix~A of Ref.~\cite{Smirnov:2012gma}).

We perform the calculation for general $R_\xi$ gauge and check that the gauge
parameter $\xi_Z$ present in the $Z$ and $\chi$ boson propagators drops out in
the result for the cross section.  In fact, it cancels between the diagrams
with top and bottom quark triangles and a neutral Goldstone boson or a $Z$
boson in the $s$ channel.  Note, that for special choices of $\xi_Z$ the
calculation can be significantly simplified. For example, in Landau gauge the
massless triangle contribution with virtual $Z$ boson
vanishes~\cite{Altenkamp:2012sx}. Note that due to Furry's theorem there is no
contribution from the vector coupling of the $Z$.  Altogether there are 
16 LO Feynman diagrams, all of them are individually finite.

We compute the LO amplitudes both in an expansion for large top quark mass
including terms up to order $1/m_t^{8}$, and without applying any
approximation and keeping the full top quark mass dependence.  In the latter
case we have reduced the tensor integrals to scalar three- and four-point
integrals which are evaluated using the {\tt LoopTools}
library~\cite{Hahn:1998yk,vanOldenborgh:1989wn}.
We want to mention that in the limit $m_t\to\infty$ the calculation is
significantly simplified.
In particular, all top quark triangle contributions with a coupling of the $Z$
boson vanish.

For the numerical results we use the following
input values~\cite{PDG}
\begin{eqnarray}
  M_Z &=& 91.1876~\mbox{GeV} \,, \nonumber\\
  M_W &=& 80.385~\mbox{GeV} \,, \nonumber\\
  M_H &=& 125~\mbox{GeV} \,, \nonumber\\
  G_\mu &=& 1.16637\cdot 10^{-5}~\mbox{GeV}^{-2} \,, \nonumber\\
  M_t &=& 173.21~\mbox{GeV} \,,
\end{eqnarray}
where $M_t$ is the top quark pole mass.
To obtain our numerical results
we follow Ref.~\cite{Altenkamp:2012sx} and use the so-called $G_\mu$
scheme where the electromagnetic coupling constant $\alpha$ and the
weak mixing angle ($s_W\equiv \sin\theta_W$) are defined via
\begin{eqnarray}
  c_W^2 &=& 1-s_W^2 \,\,=\,\, \frac{M_W^2}{M_Z^2}  \,\,\approx\,\, 0.77710
  \,,
  \nonumber\\
  \alpha &=& \frac{\sqrt{2}G_\mu M_W^2 s_W^2}{\pi} \,\,\approx\,\, 0.0075623
  \,.
\end{eqnarray}

Our default PDF set is \verb|PDF4LHC15_nlo_100_pdfas|~\cite{Butterworth:2015oua}
which we use to compute both the LO and NLO cross sections.
For the strong coupling constant we use the value provided by 
\verb|PDF4LHC15_nlo_100_pdfas| which is given by
\begin{eqnarray}
  \alpha_s(M_Z) &=& 0.118\,.
\end{eqnarray}
For the implementation of the PDFs we use version 6.1.6 of 
the {\tt LHAPDF} library~\cite{Buckley:2014ana}
(see {\tt https://lhapdf.hepforge.org/})
which also provides the running for $\alpha_s$ form $M_Z$ to the
chosen renormalization scale $\mu_R$. Our default choice
for the latter and for the factorization scale $\mu_F$
is the invariant mass of the $ZH$ system
\begin{eqnarray}
  \mu_0^2 = (p_H+p_Z)^2
  \,.
\end{eqnarray}
If not stated otherwise we choose $s_H=14$~TeV
for the hadronic center-of-mass energy.

\begin{figure}[t]
  \centering
  \includegraphics[width=\textwidth]{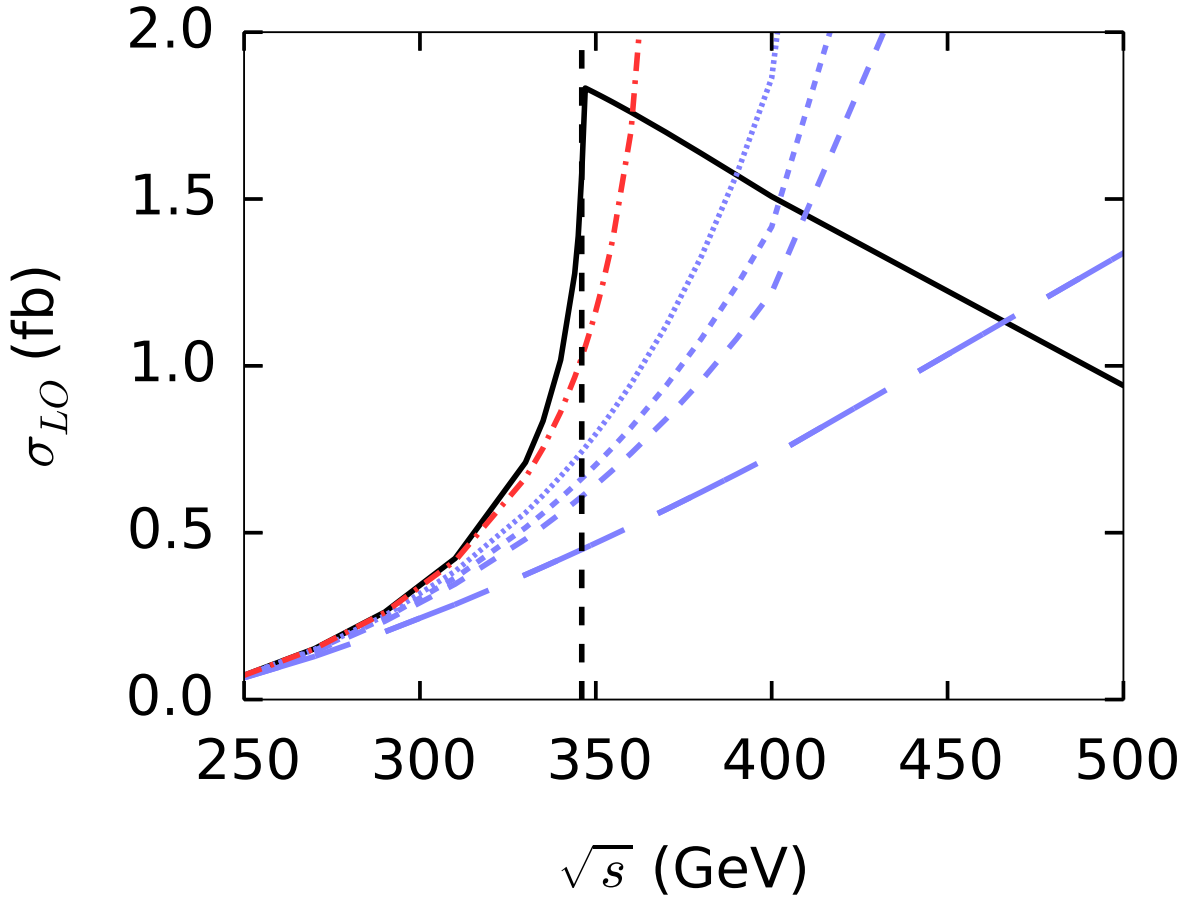}
  \caption{\label{fig::LO_exact_vs_exp}LO $gg\to ZH$
    partonic cross section as a function
    of the partonic center-of-mass energy $\sqrt{s}$. The exact result is
    shown as black solid line and the expansion terms including $1/m_t^0$,
    \ldots, $1/m_t^8$ terms (note that the $1/m_t^2$ term vanishes) are
    represented by (blue) dashed lines where shorter-dashed lines include
    higher order power corrections. The dash-dotted (red) line represents the
    $[2/2]$ Pad\'e result, see text.}
\end{figure}

In Fig.~\ref{fig::LO_exact_vs_exp} we compare the partonic cross section of
the exact (black solid line) and expanded results (blue dashed lines, see
caption for details). One observes a continuous improvement of the large-$m_t$
approximations below the top quark pair threshold which is at $\sqrt{s}\approx
346$~GeV.  However, the characteristic behaviour at threshold and the drop of
the cross section for large values of $\sqrt{s}$ cannot be reproduced. We pick
up the idea of Ref.~\cite{Campbell:2016ivq}\footnote{In contrast
  to~\cite{Campbell:2016ivq} we apply the Pad\'e approximation at the level of
  differential cross sections and not at the level of the
  amplitudes. Furthermore, we refrain from performing a conformal mapping since
  in our case the gain is marginal.}  and use the expansion terms to construct
the $[2/2]$ Pad\'e approximant, see (red) dash-dotted line in
Fig.~\ref{fig::LO_exact_vs_exp}.  One observes that the Pad\'e result
approximates reasonably well the exact curve up to $\sqrt{s}\approx 346$~GeV
which is indicated by the vertical dashed line.

\begin{figure}[t]
  \centering
  \includegraphics[width=\textwidth]{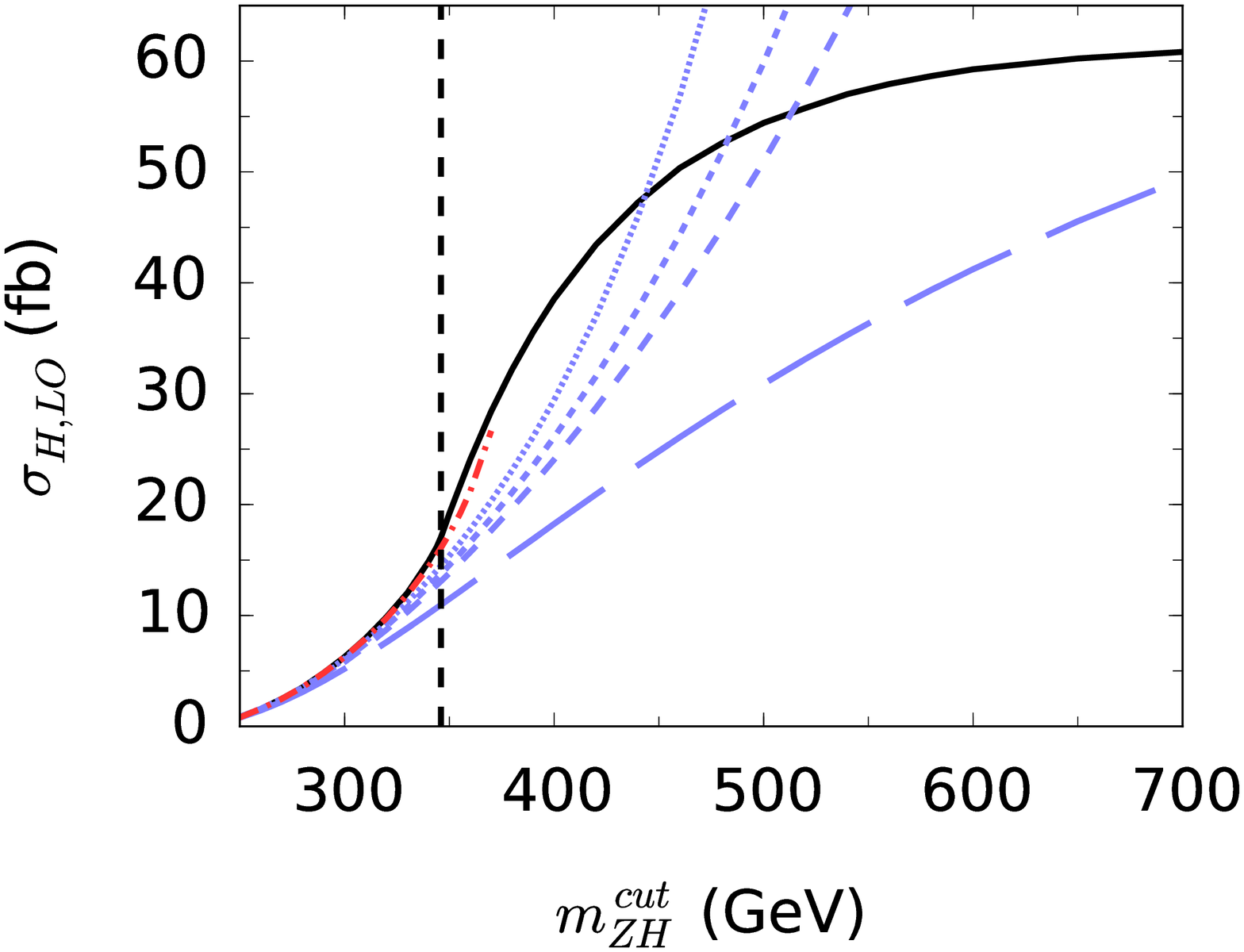}
  \caption{\label{fig::hadr_mzh_lo}Hadronic LO $gg\to ZH$
    cross section as a function of
    $m_{ZH}^{\rm cut}$, the cut on the invariant mass of the $Z$-Higgs
    system. The exact result is shown in black.  The dashed (blue) curves
    correspond to the expanded results (see caption of
    Fig.~\ref{fig::LO_exact_vs_exp} for more details) and the $[2/2]$ Pad\'e
    approximation is shown as dash-dotted (red) curve.}
\end{figure}

In Fig.~\ref{fig::hadr_mzh_lo} we show the hadronic cross section 
for $gg\to ZH$ as a
function of the cut on the invariant mass of the $Z$-Higgs system using the
same conventions as in Fig.~\ref{fig::LO_exact_vs_exp}. We observe a rapid
convergence of the $1/m_t$ expansion (blue dashed curves) for $m_{ZH}^{\rm cut}\lsim
350$~GeV and a good approximation of the exact result (solid, black) by the
Pad\'e curve (dash-dotted, red).  By construction, for large values of
$m_{ZH}^{\rm cut}$ the total cross section is reproduced.  It is interesting to note
that for $m_{ZH}^{\rm cut}\lsim 346$~GeV the cross section amounts to about
{\num a quarter} of
total cross section. For this value of $m_{ZH}^{\rm cut}$ the Pad\'e result yields
{\num 16.1}~fb which is very close to the exact result ({\num 17.0}~fb).  On
the other hand, the infinite top mass approach only gives {\num 11.0}~fb.

For a collision energy of $\sqrt{s_H}=8$~TeV we obtain for the total hadronic
cross section $\sigma_{\rm H,LO}^{\rm (exact)}={\num 16.0}$~fb which agrees
impressively well with the result obtained from the effective-theory
approximation: ${\num 15.8}$~fb.  Since the
partonic cross sections have completely different shapes (cf. solid and
long-dashed curves in Fig.~\ref{fig::LO_exact_vs_exp}) this agreement has to
be considered as accidental.  In fact, for $\sqrt{s_H}=14$~TeV we have
$\sigma_{\rm H,LO}^{\rm (exact)}={\num 61.8}$~fb whereas the
infinite-$m_t$ approximation gives ${\num 80.5}$~fb.


\section{\label{sec::part}Partonic NLO corrections}

Sample Feynman diagrams contributing to the real and virtual NLO corrections
can be found in Fig.~\ref{fig::diag}.  In our calculation we apply standard
techniques. In particular, the one- and two-loop integrals are reduced to
master integrals using the program {\tt FIRE}~\cite{Smirnov:2014hma}; the
resulting master integrals can be found in
Refs.~\cite{Gehrmann:2005pd,Ellis:2007qk}. For the isolation of the 
soft and collinear infrared divergences we follow Ref.~\cite{Frixione:1995ms}
which allows to compute differential
cross sections. Although we consider top quark mass effects we express our
final result in terms of $\alpha_s$ defined in the five-flavour theory.

We write the partonic cross section to NLO accuracy in the form
\begin{eqnarray}
  \sigma_{\rm NLO} &=& \sigma_{\rm LO}^{\rm (exact)} + \delta \sigma_{\rm NLO}^{\rm (approx)}
  + \delta \sigma_{\rm NLO}^{\rm (virt,red)}
  \,,
  \label{eq::sig_part}
\end{eqnarray}
where results for the LO cross section have already been discussed
in Section~\ref{sec::LO}.

$\delta \sigma_{\rm NLO}^{\rm (virt,red)}$ is the contribution from the
reducible diagrams where two quark triangles are connected by a gluon in the
$t$ or $u$ channel, see Fig.~\ref{fig::diag}(e) for a sample
Feynman diagram. In Ref.~\cite{Altenkamp:2012sx} the effective-theory result
for the corresponding differential cross section is given, which is obtained
by considering the interference with the LO amplitude.  We confirm the
analytic expression of~\cite{Altenkamp:2012sx} and add power-suppressed
terms up to order $1/m_t^8$. Furthermore, we have computed this contribution
exactly keeping the full top mass dependence.  For the numerical results which
we present in Section~\ref{sec::hadr_nlo} the exact expression is used.

In this section we discuss $\delta \sigma_{\rm NLO}^{\rm(approx)}$.  We define
the NLO approximation by factoring out the exact LO cross section multiplied
by the ratio of the in $1/m_t$ expanded NLO and LO contribution:
\begin{eqnarray}
  \delta \sigma_{\rm NLO}^{\rm (approx)} &=& \sigma_{\rm LO}^{\rm (exact)} 
  \,\,\frac{ \delta\sigma_{\rm NLO}^{({\rm exp}-n)} }
  { \sigma_{\rm LO}^{({\rm exp}-n)} }
  \,,
  \label{eq::sigma_NLO}
\end{eqnarray}
where ``${\rm exp}-n$'' means that the corresponding quantity
contains expansion terms up to order $1/m_t^n$.

\begin{figure}[t]
  \centering
  \includegraphics[width=\textwidth]{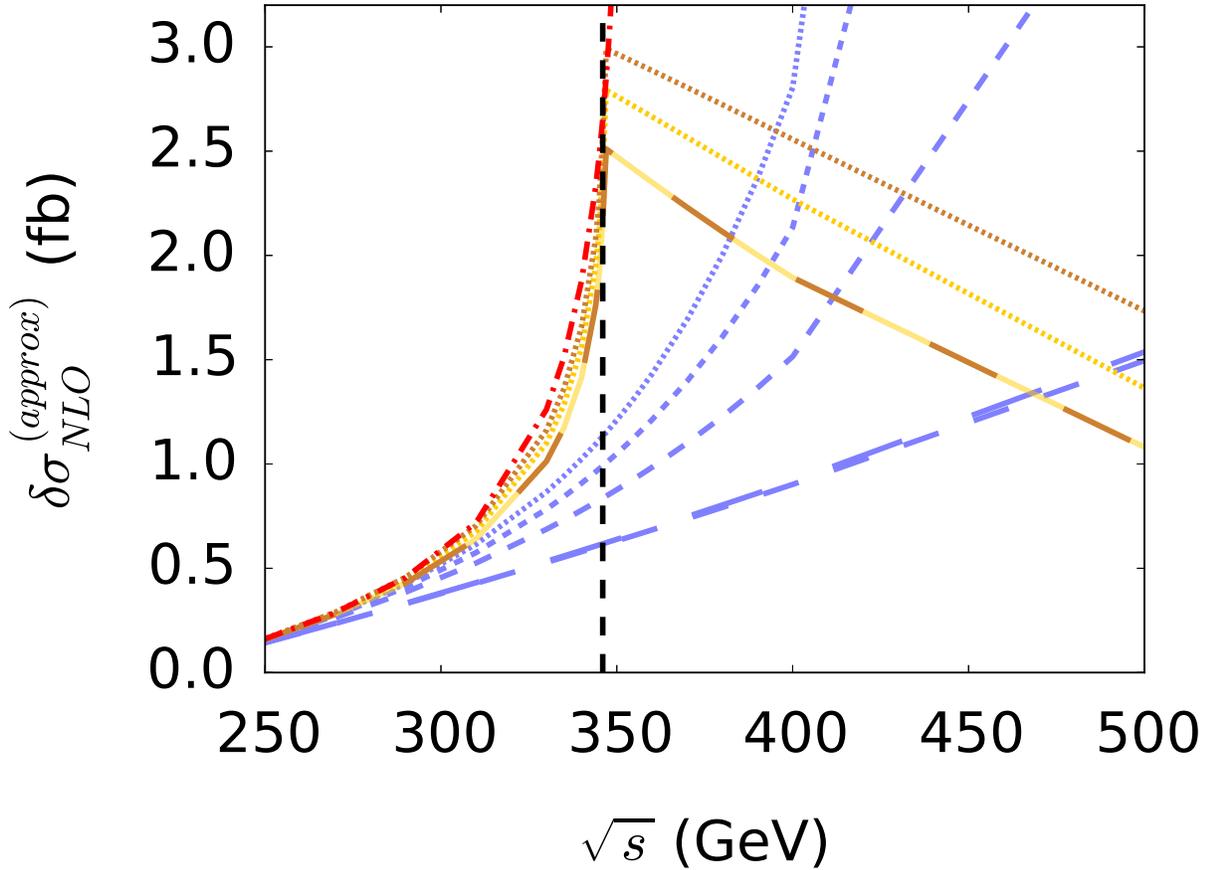}
  \caption{\label{fig::par_NLO}NLO partonic cross section as a function of
    $\sqrt{s}$. The expansion terms including $1/m_t^0$, \ldots, $1/m_t^8$
    terms are represented by (blue) dashed lines where shorter-dashed lines
    include higher order power corrections. The dash-dotted (red) line
    represents the $[2/2]$ Pad\'e result.  Approximations based on
    Eqs.~(\ref{eq::sigma_NLO}) and~(\ref{eq::nlo_fact2}) are shown as yellow
    and brown curves, respectively. In both cases we either include only the
    leading top quark mass corrections (long-dashed curves) or corrections up
    to order $1/m_t^8$ (short-dashed curves).}
\end{figure}

In Fig.~\ref{fig::par_NLO} we show as (blue) dashed lines the quantities
$\delta\sigma_{\rm NLO}^{({\rm exp}-n)}$ and as (red) dashed-dotted line the
$[2/2]$ Pad\'e approximant as a function of the partonic center-of-mass energy
$\sqrt{s}$.  We observe a similar behaviour as at LO
(cf. Fig.~\ref{fig::LO_exact_vs_exp}).  In particular, it can not be expected
that meaningful NLO approximations are obtained for large values of $\sqrt{s}$
from these expansion terms.  However, based on observations at LO we expect
that the Pad\'e result provides a reasonable approximation below
$\sqrt{s}\approx 346$~GeV.  In Fig.~\ref{fig::par_NLO} we also show as
(yellow) long- and short-dashed curves the quantity $\delta \sigma_{\rm
  NLO}^{\rm (approx)}$ with $n=0$ and $8$ (the curves for $n=2,4,6$ lie in
between and are not shown for clarity).  The shape is now dictated by the LO
cross section and has a well-behaved high-energy limit.  For $\sqrt{s} <
346$~GeV the two curves are close together, however above the top threshold
the $n=8$ curve is significantly higher.

As an alternative to Eq.~(\ref{eq::sigma_NLO}) we consider
an approach where the exact LO result is factored at the 
differential level, i.e., before the integration over
phase space. Schematically we write
\begin{eqnarray}
  \int {\rm d}{\rm PS}_2 \,
  \left|{\cal M}_{\rm LO}^{\rm (exact)}\right|^2
  \,
  \frac{ \left|{\cal M}^{({\rm exp}-n)}_{\rm NLO}\right|^2 }
  { \left|{\cal M}^{({\rm exp}-n)}_{\rm LO}\right|^2 }
  \,,
  \label{eq::nlo_fact2}
\end{eqnarray}
where ``${\rm d}{\rm PS}_2$'' indicates that we use this kind of
factorization for the two-particle phase space contributions.  The
contribution from the three-particle phase space (which is numerically
small) is added in the infinite top quark mass approximation.
The integrand of Eq.~(\ref{eq::nlo_fact2}) is better behaved than
the one for $\delta\sigma_{\rm NLO}^{({\rm exp}-n)}$ in
Eq.~(\ref{eq::sigma_NLO}), which might lead to better approximations
for the total cross section. However, below the top quark pair
threshold we only expect small differences between 
Eqs.~(\ref{eq::sigma_NLO}) and~(\ref{eq::nlo_fact2}).

Fig.~\ref{fig::par_NLO} shows $\delta \sigma_{\rm NLO}^{\rm (approx)}$ as
obtained from Eq.~(\ref{eq::nlo_fact2}) for $n=0$ and $8$ as brown dashed
lines. Note that the $n=0$ curve lies almost on top of the yellow curve (which
is based on Eq.~(\ref{eq::sigma_NLO})). This is because the two-particle phase
space contributions to the squared matrix elements are proportional to the LO
result. Moreover the three-particle contribution is small.  As before, the
$n=0$ and $n=8$ curves are close together below the top threshold and
significant deviations are observed above.


\section{\label{sec::hadr_nlo}Numerical results for hadronic cross sections}

Numerical results for the LO cross section have already been discussed
in Section~\ref{sec::LO}. At NLO we write in analogy to Eq.~(\ref{eq::sig_part})
\begin{eqnarray}
  \sigma_{\rm H,NLO} &=& 
  \sigma_{\rm H,LO}^{\rm (exact)} 
  + \delta \sigma_{\rm H,NLO}^{\rm (approx)} 
  + \delta \sigma_{\rm H,NLO}^{\rm (virt,red)}
  \,.
  \label{eq::sig_hadr}
\end{eqnarray}
For the construction of $\delta \sigma_{\rm H,NLO}^{\rm (approx)}$ we consider
three possibilities: (i) we either use the in $1/m_t$ expanded 
partonic results; (ii) we construct an approximation using 
Eq.~(\ref{eq::sigma_NLO}) (where the partonic cross sections are
replaced by their hadronic counterparts), or (iii) we utilize the differential
approach of Eq.~(\ref{eq::nlo_fact2}). The latter option is only applied to
the total cross section.

\begin{figure}[t]
  \centering
  \includegraphics[width=\textwidth]{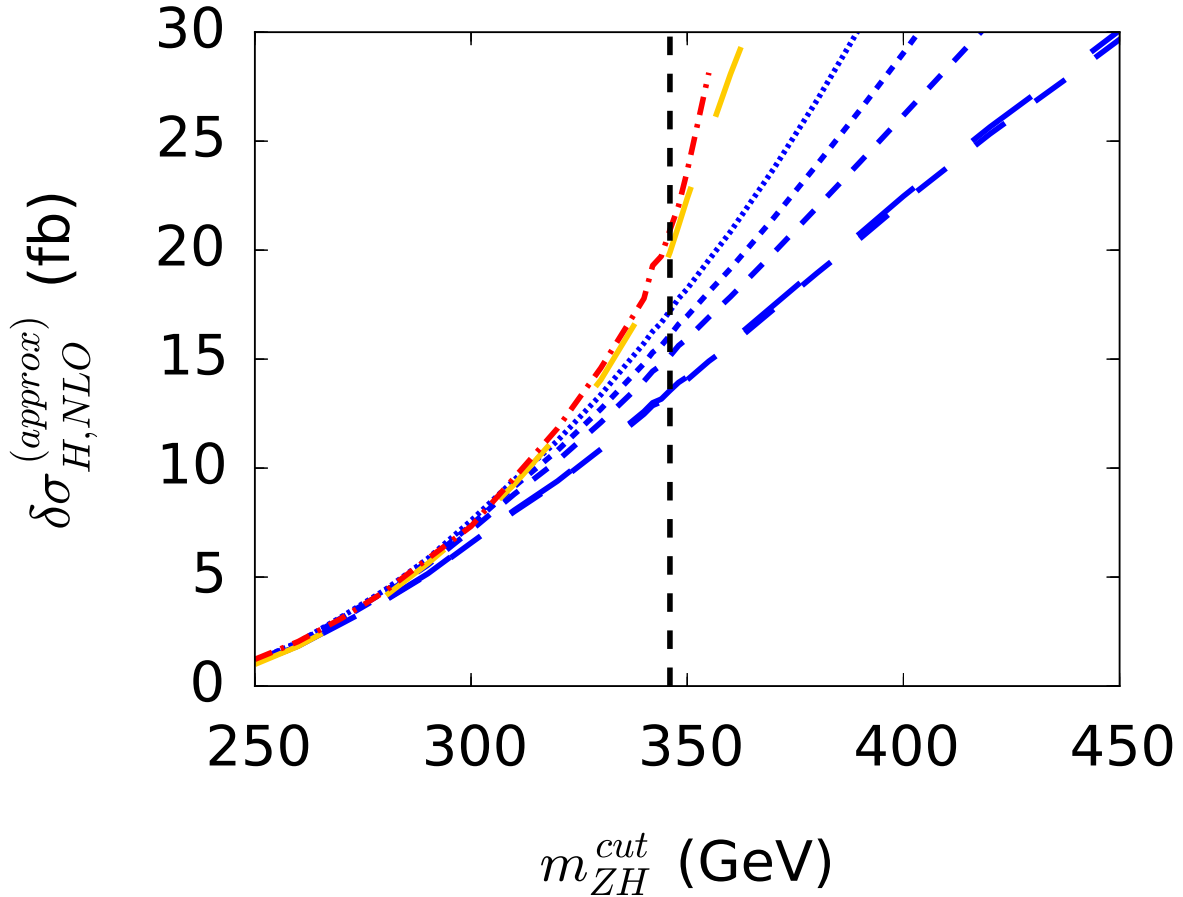}
  \caption{\label{fig::hadr_mzh_nlo}NLO contribution $\delta \sigma_{\rm
      H,NLO}^{\rm (approx)}$ to the hadronic cross section 
    as a function of $m_{ZH}^{\rm cut}$. The dashed (blue) curves contain expansion
    terms up to order $1/m_t^8$ and the dash-dotted (red) curve represents the
    Pad\'e result. The long-dashed (yellow) curve is based on
    Eq.~(\ref{eq::sigma_NLO}) with $n=0$.}
\end{figure}

Fig.~\ref{fig::hadr_mzh_nlo} shows the $m_{ZH}^{\rm cut}$ dependence of the
NLO contribution $\delta \sigma_{\rm H,NLO}^{\rm (approx)}$. We concentrate on
the region below the top quark threshold where approximations are valid. For
large values of $m_{ZH}^{\rm cut}$ one obtains the total cross section which
is briefly discussed below.  The (blue) dashed curves are obtained from the
asymptotically expanded results and the dash-dotted (red) curve is obtained
from the $[2/2]$ Pad\'e approximation.  The general picture is similar to the one at
partonic level. In particular, one observes a good convergence for
$m_{ZH}^{\rm cut}\lesssim 350$~GeV and one can expect that the Pad\'e result
provides a good approximation to the unknown exact result. Note that for
$m_{ZH}^{\rm cut}=346$~GeV the large-$m_t$ approximation gives {\num 13~fb}
whereas the Pad\'e result leads to {\num 21~fb} which corresponds to an
increase of more than 50\%.  The total cross section for $m_{ZH}^{\rm
  cut}=346$~GeV amounts to about {\num a quarter} of the total cross section computed in
the infinite top quark mass approximation (see also below).

The dashed yellow curve in Fig.~\ref{fig::hadr_mzh_nlo} is based on
Eq.~(\ref{eq::sigma_NLO}). It is obtained from the $m_{ZH}^{\rm
  cut}$-dependence of the exact LO result multiplied by the ratio of the NLO
and LO total cross sections taken in the infinite top quark mass
approximation.  Below $m_{ZH}^{\rm cut}\lesssim 350$~GeV this result and
the Pad\'e curve lie basically on top of each other.  
Very similar results are also obtained if the ratio of the $m_{ZH}^{\rm
  cut}$-dependent NLO and LO total cross sections are considered in the
effective theory limit. For reasons of clarity the corresponding curve
is not shown in Fig.~\ref{fig::hadr_mzh_nlo}.

We refrain from showing the $m_{ZH}^{\rm cut}$ dependence for $\delta
\sigma_{\rm H,NLO}^{\rm (virt,red)}$ since this contribution is numerically
small. It is negative and amounts to about {\num 1\%} of $\delta
\sigma_{\rm NLO}^{\rm (approx)}$.  However, it is included in the discussion
of the total cross section below.  Note that the infinite top quark mass
approximation of $\sigma_{\rm H,NLO}^{\rm (virt,red)}$ is off by a factor two.

\begin{table}[t]
\begin{center}
{\num
\begin{tabular}{c|c|cccc|c}
$\sqrt{s_{H}}/\text{GeV}$ & 
$\sigma_{\rm H,LO}^{\rm (exact)}$ & 
$\sigma_{\rm H,NLO}^{\rm (exp-0)}$ & 
$\sigma_{\rm H,NLO}^{\rm (re-scale)}$ & 
$\sigma_{\rm H,NLO}^{\rm (diff)}$ & 
$\sigma_{\rm H,NLO}^{[2/2]}$ & 
NLO scale variation\\
\hline
7  & 11.2 & 23.9  & 24.9  & 26.1 & 26.5  & ${}^{+21\%}_{-21\%}$\\
8  & 16.0 & 35.2  & 35.4  & 37.2 & 38.8  & ${}^{+20\%}_{-20\%}$\\
13 & 52.4 & 129 & 113 & 121 & 140 & ${}^{+14\%}_{-17\%}$\\
14 & 61.8 & 155 & 133 & 142 & 168 & ${}^{+13\%}_{-16\%}$

\end{tabular}
}
\caption{\label{tab:totalCS} LO and NLO results for the total cross section in
  fb.  In columns 3 to 6 the following NLO contributions are added to the
  exact LO result: infinite top mass approximation ($\sigma_{\rm H,NLO}^{\rm
    (exp-0)}$), re-scaled NLO contribution based on Eq.~(\ref{eq::sigma_NLO})
  ($\sigma_{\rm H,NLO}^{\rm (re-scale)}$), re-scaled NLO contribution based on
  Eq.~(\ref{eq::nlo_fact2}) ($\sigma_{\rm H,NLO}^{\rm (diff)}$), and the
  approximation where below the top threshold the [2/2]-Pad\'e result and
  above the infinite top mass approximation is used ($\sigma_{\rm
    H,NLO}^{[2/2]}$).  The last column gives the scale uncertainties for
  $\sigma_{\rm H,NLO}^{[2/2]}$ where $\mu_F=\mu_R$ is varied by
  $\mu_F/\mu_0\in[1/3,3]$.  The NLO cross sections contain $\sigma_{\rm
    H,NLO}^{\rm (virt,red)}$.}
\end{center}
\end{table}

Table \ref{tab:totalCS} shows the values for the total cross section at LO and
for four possible approximations at NLO, see caption for details.  The first
three approximations treat the top quark as infinitely heavy, whereas the
third one incorporates the heavy quark effects considered earlier in the form
of a $[2/2]$ Pad\'e approximation. Note, that we only consider finite top mass
corrections for $\sqrt{s}<346~\text{GeV}$. For higher values of $s$ the
infinite top mass limit is applied. One observes, that the finite top mass
corrections shift the total cross section upwards, however, the size is well
within the scale uncertainties which are shown for $\sigma_{\rm
  H,NLO}^{[2/2]}$ in the last column. Similar uncertainties are also obtained
for the other approximations.

The numerical results discussed in this section and in Section~\ref{sec::part}
have been obtained with the help of the program {\tt ggzh} which can be
downloaded from~\cite{ggzh}. A brief description
of {\tt ggzh} can be found in the appendix.
{\tt ggzh} can be used to reproduce the numerical results of Ref.~\cite{Altenkamp:2012sx}.



\section{\label{sec::concl}Conclusions}

The associated production of a Higgs and $Z$ boson is a promising channel in
view of the determination of the Higgs boson couplings, in particular the
Yukawa coupling to bottom quarks.  We compute top quark mass effects to the
loop-induced process $gg\to ZH$ at NLO in QCD by expanding the Feynman
amplitudes in the limit of large top quark mass. Our leading term reproduces
the results of Ref.~\cite{Altenkamp:2012sx}. It is not expected that the top
quark suppressed terms provide a good approximation for large partonic
center-of-mass energies. However, we can show that below the production
threshold of two top quarks, say for $\sqrt{s}\lesssim350$~GeV, the
$1/m_t$-expansion shows a good convergence at NLO. This is strongly
supported by the good agreement of the re-scaled NLO approximation using the
exact LO cross section and the $[2/2]$ Pad\'e approximation constructed from
expansion terms up to $1/m_t^8$.  Thus, the corrections computed in this paper
provide a good approximation to the $m_{ZH}$ distributions below
$\sqrt{s}\lesssim350$~GeV. This region covers about {\num 25\%} of the total
cross section.  Furthermore, the top mass corrections in this region
constitute an important cross check once the exact calculation of the NLO
corrections to $gg\to ZH$ is available.  The numerical results presented in
this work can be reproduced with the program {\tt ggzh} which is publicly
available from~\cite{ggzh}.


 
\section*{Acknowledgements}

We are thankful to Kirill Melnikov for enlightening discussions and to Lorenzo
Tancredi for providing analytic results for the amplitude $Z\rightarrow ggg$
from Ref.~\cite{Gehrmann:2013vga} which we could compare to our results.  We
thank Robert Harlander for helping with the comparison to {\tt
  vh@nnlo}~\cite{Brein:2012ne}.  This work is supported by the Deutsche
Forschungsgemeinschaft through grant STE~945/2-1.



\begin{appendix}

\section{Brief description of {\tt ggzh}}

Together with this paper we also publish the program {\tt ggzh}
which can be downloaded from~\cite{ggzh}.
{\tt ggzh} includes all contributions to the process $gg \to ZH$
which are discussed in this paper.

{\tt ggzh} is written in {\tt C++}.  Before compilation it is necessary to
install the libraries CUBA~\cite{Hahn:2004fe}, {\tt
  LoopTools}~\cite{Hahn:1998yk,vanOldenborgh:1989wn} and {\tt gsl}~\cite{gsl}.
The corresponding paths should be inserted in the file {\tt Makefile.local}.
Afterwards, {\tt make} starts the compilation.

The input file {\tt xsection.cfg} defines the channels which shall
be considered. Furthermore, one has to decide whether the partonic or hadronic
cross section is considered, which pdf set is used and whether the 
sum of the considered channels is computed or not. Thus, 
{\tt xsection.cfg} typically looks as follows
\begin{verbatim}
  active channels: {LO_exact,LO_0}
  pdf set: PDF4LHC15_nlo_100_pdfas
  hadronic: true
  sum channels: false
\end{verbatim}
{\tt ggzh} outputs partonic cross sections in case \verb|hadronic: false|
is chosen. In the sample file the exact LO cross section and the
effective-theory result including $1/m_t^0$ terms is computed.
Further available channels are
\verb|LO_<i>| with $i=2,4,6,8$ for the $1/m_t^i$ contribution
and \verb|pade22_LO| for the $[2/2]$ Pad\'e approximation of the LO
cross section.
The $1/m_t^i$ contribution to $\delta \sigma_{\rm NLO}^{\rm (approx)}$
is obtained by summing the channels
\verb|NLO_phase2_<i>|, \verb|NLO_phase2eta_<i>| and \verb|NLO_phase3_<i>|
($i=0,2,4,6,8$) and $\delta \sigma_{\rm NLO}^{\rm (virt,red)}$
is implemented in \verb|NLO_reducible_exact|.

Results based on the differential factorization
of Eq.~(\ref{eq::nlo_fact2}) can be obtained
via the channels \verb|NLO_differential_phase2| and
\verb|NLO_differential_phase2eta|
(remember that Eq.~(\ref{eq::nlo_fact2}) is only applied to two-particle phase
space contributions). The parameter \verb|diff_order| in the input file
{\tt params.cfg} specifies the expansion depth used for the LO and NLO
expressions in~(\ref{eq::nlo_fact2}).

The second input file {\tt params.cfg} contains the values for
the various input parameters needed for the calculation. It
overwrites the default values which are given in {\tt params.def}
together with a brief description of the meaning.
The package comes with template files which clarify the syntax.

{\tt ggzh} is launched by simply calling the executable in the shell
\begin{verbatim}
  > ./ggzh
\end{verbatim}
All input parameter are repeated in the output and
the results for the individual channels is given in the form
\begin{verbatim}  
Calculating hadronic cross-section for channel "LO_exact".
Integrating (Vegas) ...
Number of integrand evaluations: 1050000
Integration time: 49s. Per iteration: 0.04697ms
Result [1/(GeV)^2]: 1.5875696750976581e-10
Error  [1/(GeV)^2]: 7.5506640742330014e-13
Result [fbarn]:     61.816682911840118
Error  [fbarn]:     0.29400725786852289
Relative error:     0.0047561150812284996
Chi^2 Probability:  0.18458145471778875
#points dropped:    0

Calculating hadronic cross-section for channel "LO_0".
Integrating (Vegas) ...
Number of integrand evaluations: 1050000
Integration time: 2s. Per iteration: 0.002444ms
Result [1/(GeV)^2]: 2.0665863953725464e-10
Error  [1/(GeV)^2]: 1.3007066760509459e-13
Result [fbarn]:     80.468604254996833
Error  [fbarn]:     0.050646830445289774
Relative error:     0.00062939864452967408
Chi^2 Probability:  6.004534622038346e-12
#points dropped:    0
\end{verbatim}

Besides the total cross section it is also possible to introduce a cut
on the invariant mass $m_{ZH}$ which is switched on with 
\verb|use_inv_mass_cutoff: 1| in the file
{\tt params.cfg}. The numerical values for the cut is specified with
\verb|inv_mass_cutoff: <m_ZH-value>|.

With the help of \verb|use_mt_threshold: 1| one switches on the possibility to
use the infinite top mass approximation above the value for $\sqrt{s}$ given
by \verb|mt_threshold: <mtthr-value>|.

{\tt ggzh} contains the option to vary
$\mu_R$ and $\mu_F$ independently. Furthermore, it
is possible to choose fixed scales (e.g. $\mu_R=M_H$ or $\mu_R=m_t$)
or identify the scales to the partonic center-of-mass energy.

\end{appendix}


\end{document}